\documentclass{article}
\usepackage[utf8]{inputenc}
\usepackage{amsmath,amssymb}
\usepackage{bm, bbm, blkarray, adjustbox}
\usepackage{amsfonts}
\usepackage{enumitem}
\usepackage[margin=1in]{geometry}
\usepackage{natbib}
\usepackage{hyperref}

\usepackage{graphicx}
\usepackage{soul}
\usepackage{setspace}
\usepackage{listings}
\usepackage{color,soul}
\usepackage{authblk}
\usepackage{booktabs}

\renewcommand{\d}[1]{\ensuremath{\operatorname{d}\!{#1}}}

\title{A note on promotion time cure models\\ with a new biological consideration}

\author[1]{Zhi Zhao}
\author[1,2]{Fatih K\i z\i laslan}

\affil[1]{Department of Biostatistics, University of Oslo\\
P.O.Box 1122 Blindern, 0317 Oslo, Norway}
\affil[2]{Department for Method Development and Analytics, \\
Norwegian Institute of Public Health, Oslo, Norway}

\begin{document}
\date{}

\maketitle

\begin{abstract}
We introduce a generalized promotion time cure model motivated by a new biological consideration. 
The new approach is flexible to model heterogeneous survival data, in particular for addressing intra-sample 
heterogeneity. 
We also indicate that the new approach is suited to model a series or parallel system consisting of multiple subsystems in reliability analysis.
\end{abstract}



\section{Introduction}\label{sec:intro}

The promotion time cure model (PTCM) is one of the most important models in survival analysis, but it has not yet been studied much in the literature \citep{Amico2018}. 
The PTCM is constructed by motivating biological considerations, which assumes that after initial treatment the time to recurrence of cancer is the result of a latent process of the residual tumor cells (i.e. clonogenic cells) propagating into a newly detectable tumor \citep{Yakovlev1996}. 
As shown in \cite{Yakovlev1996} and \cite{Chen1999}, the construction of the PTCM dependents on a latent variable $N$ that is the number of clonogenic cells left active in a patient after initial treatment. 
Assume that $N$ is Poisson distributed with mean $\theta>0$, i.e. $\mathbb P(N=k) = \theta^k e^{-\theta} / k!$, $k\in \mathbb N$.
Let another latent variable $Z_{j}$ ($j=1,...,N$) be the random time for the $j$-th clonogenic cell to produce a detectable tumor mass. 
Given $N$, the variables $Z_{j}$ are independently and identically distributed with cumulative distribution function (cdf) $F(t) = 1-S(t)$. 
Here $F(t)$ is the promotion time distribution of any clonogenic cell and $S(t)$ is its corresponding survival function. 
The time to tumor recurrence can be defined by the random variable $T=\min_{0\leq j\leq N} \{Z_{j}\}$, i.e. tumor recurrence when the one of the clonogenic cells becomes activated, where $\mathbb P(Z_{0}=\infty )=1$ (i.e. no tumor recurrence in a finite time). 
Note that the time to tumor recurrence $T$ of a patient is observable, but $N$ and $Z_{j}$ are unobservable latent variables. 
The survival function of the population is the probability of no newly detectable tumor by time $t$
given by 
{\small
\begin{align*}
S_{pop}(t) 
&= \mathbb P(N=0)+ \mathbb P(Z_{1}>t,....,Z_{N}>t,N> 0)  \\
&= e^{-\theta }+\sum_{k=1}^{\infty } \mathbb P(Z_{1}>t,....,Z_{N=k}>t) \mathbb P(N=k)  \\
&= e^{-\theta }+\sum_{k=1}^{\infty }S(t)^{k}\frac{\theta ^{k}e^{-\theta }}{k!} 
= e^{-\theta \left( 1-S(t)\right) } = e^{-\theta F(t)}. \tag{1}\label{S_pop}
\end{align*}
}
Covariates can be introduced in the parameter $\theta$ and may also be introduced in the proper baseline distribution function $F(t)$. 
However, it is not clear how to include clonogenic cell data information in the PTCM (\ref{S_pop}) for better progression-free survival prediction or potentially for the identification of subclonal driver genes predictive of survival. 
Motivated by this, we propose the following generalized promotion time cure model (GPTCM) to integrate multiscale data, i.e. cancer patient data on multiple biological scales: individual-level survival data, cellular-level cell type proportions data, and subcellular-level cell-type-specific genetic variables.

\vspace{-5mm}
\section{The generalized promotion time cure model (GPTCM)}\label{sec:method}

\subsection{Formulation}\label{subsec:formulation}

In cancer cell biology, the tumor might contain a mixture of cell subtypes, for example, invasive tumor cells, non-invasive tumor cells and stromal cells \citep{Trapnell2015}. 
Motivated by the classical PTCM (\ref{S_pop}), we assume that all tumor cells are composed of multiple clonogenic cell groups (i.e. tumor cell subtypes or subclones). 
Suppose a patient after an initial treatment has the total number of tumor cells $N = \sum_{l=1}^L N_l$, $L \ge 2$, where $N_l$ is the number of $l$-th cluster of cells (e.g. tumor cell subtype). 
%
%
Similar to the PTCM, let the $l$-th cluster have multivariate random times for $N_l$ clonogenic cells propagating into a newly detectable tumor:
$$
\bm W_l = \left(Z_{\sum_{j=1}^{l-1}N_{j-1}+1}, ..., Z_{\sum_{j=1}^{l-1}N_{j-1}+N_l}\right),
$$ 
where $l\in \{1,...,L\}$ and $N_0=0$. 
For the $N_l$ homogeneous cells in the $l$-th cluster, we assume cluster-specific promotion time distribution $F_l(t)=1-S_l(t)$, and then we have 
$$
\mathbb P(\min\{\bm W_l\} >t) = S_l(t)^{N_l}.
$$
The time to tumor recurrence can be defined as 
$T=\min \{\min\{\bm W_1\},...,\min\{\bm W_L\}\}$. 
Then the survival function for the population is given by
%
\begin{align*}
  S_{pop}(t) &= 
  \mathbb P(\text{no cancer by time }t) \\
  &= \mathbb P(N=0) + \mathbb P\left(\min\{\min\{\bm W_1\},...,\min\{\bm W_L\}\} >t, N>0\right) \\
  &= \mathbb P(N=0) + \mathbb P(\min\{\bm W_1\} >t, ..., \min\{\bm W_L\}>t, N>0) \\
  &= \mathbb P(N=0) + \sum_{k=1}^\infty\mathbb P\left(\min\{\bm W_1\} >t, ..., \min\{\bm W_L\}>t | N=k\right) \cdot \mathbb P(N=k)
  . \tag{2}\label{S_pop2}
\end{align*}
%
To compute the second term in (\ref{S_pop2}), 
we use the multinomial theorem and the multinomial-Poisson transformation \citep{Brookmeyer1989}. 
With the multinomial theorem, we can consider all configurations of $(N_1,...,N_L)$ such that their sum is $k$. 
If $N_l$'s are independent Poisson random variables with mean $\theta_l$ denoted as $\mathcal Pois(N_l; \theta_l)$, ($l=1,...,L$), by the multinomial-Poisson transformation, the unconditional joint distribution of $(N_1,...,N_L)$ can be factorized into the product of a Poisson distribution and a multinomial distribution. 
The multinomial distribution is 
$\mathbb P(N_1,...,N_L|N=k) =: \mathcal Mult(k;\bm p)$, 
where 
$\bm p=(p_1,...,p_L)$, 
$p_l=\theta_l/\theta, \theta=\sum_l\theta_l$. 
Then we obtain 
{\small
\begin{align*}
  &S_{>0}(t) := 
  \sum_{k=1}^\infty\mathbb P\left(\min\{\bm W_1\} >t, ..., \min\{\bm W_L\}>t | N=k\right) \cdot \mathbb P(N=k)\\
  &= \sum_{k=1}^\infty\sum_{\substack{\text{All config.}\\(N_1,...,N_L)\\ \text{ with sum }k}}\mathbb P\left(\min\{\bm W_1\} >t,..., \min\{\bm W_L\}>t |N=k, N_1,...,N_L\right) \cdot \mathbb P(N=k,N_1,...,N_L)\\ 
  &= \sum_{k=1}^\infty\sum_{\substack{\text{All config.}\\(N_1,...,N_L)\\ \text{ with sum }k}} \mathbb P\left(\min\{\bm W_1\} >t, ..., \min\{\bm W_L\}>t | N=k,N_1,...,N_L\right) 
  \cdot \mathbb P(N=k)\mathbb P(N_1,...,N_L| N=k)\\ 
  &= \sum_{k=1}^\infty\sum_{\substack{\text{All config.}\\(N_1,...,N_L)\\ \text{ with sum }k}}\mathbb P\left(\min\{\bm W_1\} >t, ..., \min\{\bm W_L\}>t | N=k,N_1,...,N_L\right) \cdot \mathcal Pois(k;\theta)\mathcal Mult(k;\bm p)\\ 
  &= \sum_{k=1}^\infty\sum_{\substack{\text{All config.}\\(N_1,...,N_L)\\ \text{ with sum }k}} \prod_{l=1}^L \mathbb P(\min\{\bm W_l\} >t | N_l) \cdot \mathcal Pois(k;\theta)\mathcal Mult(k;\bm p)\\ 
  &= \sum_{k=1}^\infty\sum_{\substack{\text{All config.}\\(N_1,...,N_L)\\ \text{ with sum }k}} \prod_{l=1}^L S_l(t)^{N_l} \cdot \mathcal Pois(k;\theta)\mathcal Mult(k;\bm p)\\ 
  &= \sum_{k=1}^\infty\sum_{\substack{\text{All config.}\\(N_1,...,N_L)\\ \text{ with sum }k}} \prod_{l=1}^L S_l(t)^{N_l} \cdot \frac{\theta^k e^{-\theta}}{k!} \cdot \frac{k!}{N_1!... N_L!}p_1^{N_1}... p_L^{N_L}\\ 
  &= \sum_{k=1}^\infty\frac{\theta^k e^{-\theta}}{k!} \sum_{\substack{\text{All config.}\\(N_1,...,N_L)\\ \text{ with sum }k}} \frac{k!}{N_1!... N_L!}\{p_1S_1(t)\}^{N_1}... \{p_LS_L(t)\}^{N_L} \\ 
  &= \sum_{k=1}^\infty\frac{\theta^k e^{-\theta}}{k!} \{p_1S_1(t) +... + p_LS_L(t)\}^k \\
  &= \left\{e^{\theta\sum_{l=1}^Lp_lS_l(t)} - 1 \right\} e^{-\theta}
  . 
\end{align*}
}
%
Finally, the population survival function is
\begin{align*}
  S_{pop}(t) &= \mathbb P(N=0) + S_{>0}(t) \\
  &= e^{-\theta} + \left\{e^{\theta\sum_{l=1}^Lp_lS_l(t)} - 1 \right\}e^{-\theta} \\
  &= e^{-\theta\left\{1 - \sum_{l=1}^Lp_lS_l(t) \right\}}
  . \tag{3}\label{S_pop_new}
\end{align*}
%
Let $F(t) = 1-\sum_{l=1}^Lp_lS_l(t)$, and then $S_{pop}(t)=e^{-\theta F(t)}$. 
This means that if $S(t)=S_l(t)$, $\forall l\in \{1,...,L\}$ (i.e. no different types of cells), the population survival function (\ref{S_pop_new}) is degenerated into PTCM (\ref{S_pop}), 
so the new model is named as generalized promotion time cure model (GPTCM). 
\smallskip\\
{\noindent\bf Remark 1.} Note that the proportions $(p_1,...,p_L)$ in the GPTCM (\ref{S_pop_new}) are patients' cancer cell proportions data (i.e. $n$-by-$L$ data matrix collected from $n$ patients) not simple weight parameters. 
Therefore, the GPTCM can integrate multiscale data [i.e. individual-level survival data, cellular-level cell type proportions data, and subcellular-level cell-type-specific molecular and genomic data (see Remark 4 below)] for joint modeling.
\smallskip\\
{\noindent\bf Remark 2.} The GPTCM is similar to the general class of PTCM with Equation 2 in \cite{Gomez2023} whose population survival function was derived based on a compound Poisson distribution for the total number of clonogenic cells. 
But they assumed a common promotion time distribution for all cells, i.e. without distinguishing heterogeneous cells. 
\cite{Gomez2023} also fixed the number of cells to be 1, 2 or $\infty$ regardless of the exact tumor cells in patients. 
\smallskip\\
{\noindent\bf Remark 3.} The GPTCM is also similar to a mixtures-of-experts model for survival analysis \citep{Rosen1999} or a generalized Weibull mixture model 
for reliability analysis \citep{Jiang1995}. 
But the (generalized) mixture models are to model inter-sample heterogeneity, assuming every sample is from one of the mixture clusters. 
In contrast, our GPTCM is to model intra-sample heterogeneity, since every sample/patient has data for multiple clusters of $p_lS_l(t), l\in \{1,...,L\}$. 
In fact, the formulation of a mixture model like 
$
S_{pop}(t)=\mathbb P(N=0) + \sum_{l=1}^Lp_lS_l(t)
$
is not biologically meaningful in the situation that the time to recurrence is the result of a latent process for cancer recurrence, see Appendix \ref{sec:appendixA}. 
\smallskip\\
{\noindent\bf Remark 4.} Similar to the PTCM, the GPTCM can introduce covariates $X$ through the Poisson rate parameter $\theta$, e.g. $\theta=\exp(\xi_0+ X\xi)$. 
Benefiting from the mixture part in the GPTCM, cluster-specific covariates (e.g. genetic variables from each tumor cell subtype) $X_l$ can be introduced in $S_l(t)$. 
For example, using a log-linear model to capture the mean survival time, i.e. $\log\mu_l= X_l\beta_l, l\in \{1,...,L\}$, where $\mu_l$ is the mean of the Weibull distribution 
$
S_l(t) = \exp\{ -(t / \lambda_l)^\kappa \}, 
\lambda_l = \mu_l / \Gamma(1+1/\kappa), 
\kappa \in \mathbb R_+, 
$
and $\Gamma(\cdot)$ is the gamma function. 
The modeling of tumor cell-type-specific genes has the potential to identify cell-type-specific drivers for cancer prognosis, and ultimately improve individualized cancer diagnosis and personalized cancer therapies. 
Furthermore, if we assume randomness in the proportions data (e.g. following Dirichlet distribution), any covariate may also be introduced to model the compositional data of cell proportions \citep{Greenacre2021, Mangiola2023}. 
%
\smallskip\\
{\noindent\bf Remark 5.} Identifiability is an important issue in the estimation of cure models. 
The GPTCM is identifiable when $\theta=\exp(\xi_0+ X\xi)$ and $\lim_{t\to \infty}S_l(t)=0$ ($\forall l \in\{1,...,L\}$) according to Proposition 7 in \cite{Hanin2014}. 
In a finite mixture model $\sum_{l=1}^Lp_lS_l(t)$, the label switching problem is a common identifiability issue, since there is no prior information to distinguish between the clusters of the mixture. 
However, in the applications of single-cell data, cell types can be predefined based on cell biology, and single-cell sequencing data usually result in well estimated cell type proportions. 
As mentioned in Remark 1, the proportions $(p_1,...,p_L)$ in the GPTCM are cancer cell proportions data collected from patients rather than weight parameters, so the label switching is irrelevant.



\subsection{Connection to last-activation scheme and reliability analysis} 

The PTCM is also referred to as the first-activation scheme \citep{Cooner_2007}. 
When all clonogenic cells are homogeneous (i.e. no different types of cells) and the time to tumor recurrence is when the last clonogenic cell becomes activated (i.e. $T=\max_{0\leq j\leq N} \{Z_{j}\}$), \cite{Cooner_2007} referred this as the last-activation scheme and its corresponding population survival function is 
$
1+ e^{-\theta}(1-e^{\theta F(t)})
$. 
Similar to the last-activation scheme, the GPTCM can be extended for the recurrence to be observed when the last class of clonogenic cells becomes activated. 
Then the time to tumor recurrence can be defined as $T=\max \{\max\{\bm W_1\},...,\max\{\bm W_L\}\}$ and the population survival function is given by (see Appendix \ref{sec:appendixB} for details)
\begin{equation}
 \tilde S_{pop}(t) = 1 + e^{-\theta} - e^{-\theta\sum_{l=1}^Lp_lS_l(t)}.
 \tag{4}\label{S_last_acti}
\end{equation}
Here $\tilde S_{pop}(0)$ is improper with $\tilde S_{pop}(0) = 1$ and cure rate $\tilde S_{pop}(\infty) = e^{-\theta}$. 

\begin{figure}
\centering
    \includegraphics[height=0.47\textwidth]{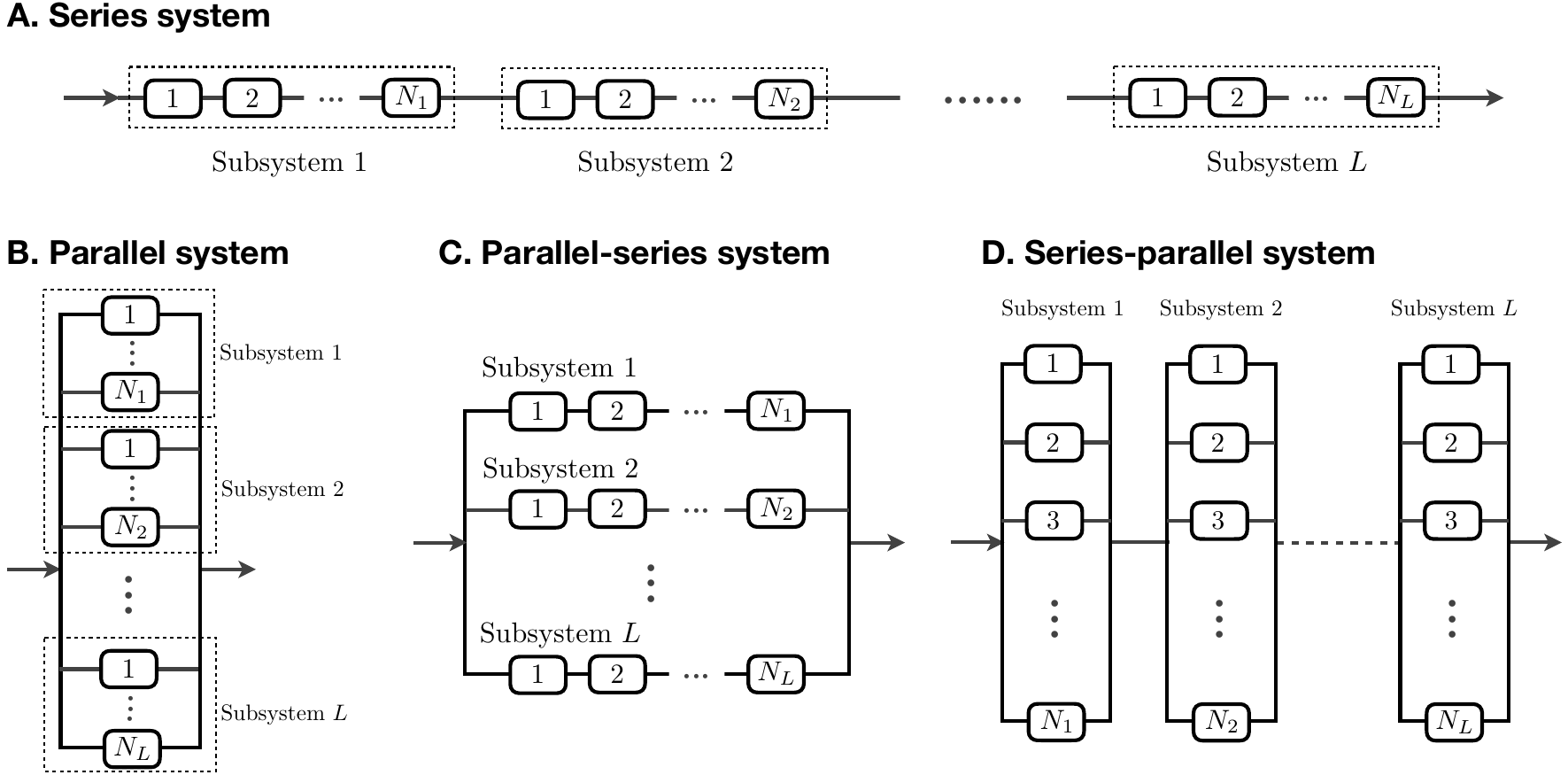}
    \caption{\it Illustration of series system, parallel system, parallel-series system and series-parallel system.}
    \label{fig:reliability}
\end{figure}

From the perspective of system reliability, the PTCM can be interpreted as analogous to a series system with a random number of units under random shock \citep{Cha2018}. 
In such a system, failure occurs as soon as one unit fails, making the PTCM conceptually similar to a reliability structure where the weakest link dictates the overall system failure. 
Our proposed GPTCM can be suited to model a system consisting of multiple heterogeneous subsystems (Fig. \ref{fig:reliability}A), as discussed in \cite{Wei2023}, which investigates the reliability of the time until one critical subsystem fails.

In reliability engineering, a natural extension to the series system is the latent parallel system model, in which failure occurs only after all latent factors have been activated, known as the last-activation scheme defined in \cite{Cooner_2007}. 
It represents a contrasting mechanism where the survival time depends on the simultaneous activation of multiple latent processes, rather than being dictated by the earliest activation. 
Therefore, the last-activation scheme model (\ref{S_last_acti}) can be used for a parallel system with multiple subsystems (Fig. \ref{fig:reliability}B). 
The cure rate $\tilde S_{pop}(\infty) = e^{-\theta}$ means that a harmful event does not result in an ultimate system failure. 
Further extensions can be for a parallel-series system (Fig. \ref{fig:reliability}C) with the failure time $T=\max \{\min\{\bm W_1\},...,\min\{\bm W_L\}\}$, i.e. the failure occurs when all of the parallel subsystems fail, 
and can also be for a series-parallel system (Fig. \ref{fig:reliability}D) with the failure time $T=\min \{\max\{\bm W_1\},...,\max\{\bm W_L\}\}$, i.e. the failure occurs when one of the parallel subsystems fails. 

\subsection{Statistical characteristics of the GPTCM}

The GPTCM (for the first-activation scheme) and the classical PTCM have similar statistical properties. 
For example, both the PTCM and GPTCM do not have proper survival functions, since their cure fraction is $S_{pop}(\infty) = e^{-\theta} >0$. 
The survival function of the noncured population of the GPTCM is a proper survival function, i.e. $S^*(t)=S_{>0}(t)/(1-e^{-\theta})$, $S^*(0) = 1$ and $S^*(\infty) = 0$. 
Assuming all covariates $X$ are time-independent, 
the population probability density function (pdf) of the GPTCM is given by
\begin{align*}
  f_{pop}(t|X) 
  = -\frac{\d S_{pop}(t|X) }{\d t} 
  = \theta f(t) e^{-\theta F(t)}, 
\end{align*}
where $F(t) = 1-\sum_{l=1}^Lp_lS_l(t)=\sum_{l=1}^Lp_lF_l(t)$, 
$f(t) = (\d /\d t) F(t)=\sum_{l=1}^L p_l f_l(t)$, 
and $f_l(t)$ and $F_l(t)$ are the cluster-specific promotion time pdf and cdf, respectively. 
Note that here $f_{pop}(t|X)$ is not a proper pdf, since $S_{pop}(t|X)$ is not a proper survival function.


The hazard functions of the entire population and the noncured population of the GPTCM are
\begin{align*}
h_{pop}(t|X) &= \frac{-\d S_{pop}(t|X) / \d t}{S_{pop}(t|X)} 
= \theta f(t) = \theta\sum_{l=1}^L p_l f_l(t), \\
h^*(t|X) &= \frac{-\d S^*(t|X) / \d t}{S^*(t|X)}
= \left\{1-e^{-\theta \sum_{l=1}^L p_l S_l(t) }\right\}^{-1} \theta \sum_{l=1}^L p_l f_l(t).
\end{align*}
\begin{figure}
\centering
    \includegraphics[height=0.3\textwidth]{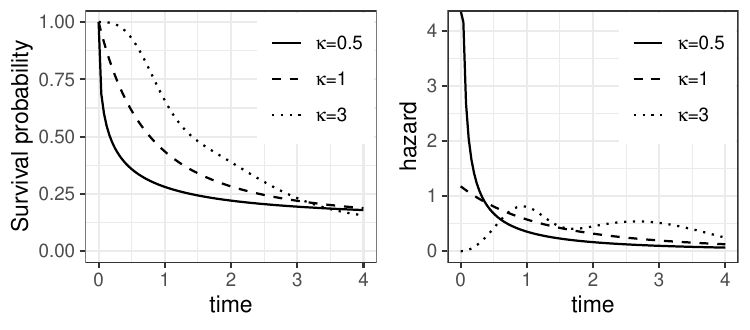}
    \caption{\it Survival and hazard plots of the GPTCM with two clusters. Weibull distributions with cluster-specific scale parameters and common shape parameter are as an example: $S_l(t)=e^{-(t/\lambda_l)^\kappa}$, $\lambda_l = \frac{\mu_l}{\Gamma(1+1/\kappa)}$, $l\in \{1,2\}$, $(\log\mu_1,\log\mu_2)=(-0.1, 1)$, $(p_1,p_2)=(0.3, 0.7)$, $\theta=2$, and the Weibull's shape parameter $\kappa=0.5$ (solid line), $\kappa=1$ (dashed line) or $\kappa=3$ (dotted line).}
    \label{fig:survival}
\end{figure}
Similar to \cite{Chen1999}, we also have 
$$ 
h_{pop}(t|X) = h^*(t|X) \left\{ 1-e^{-\theta \sum_{l=1}^L p_l S_l(t) } \right\} \le h^*(t|X), 
$$ 
i.e. the hazard function of the noncured samples is greater than a sample selected from the entire population. 
We can also obtain the population cumulative hazard function $H_{pop}(t)=\int_0^t h_{pop}(s)\d s=\theta\sum_{l=1}^L p_l F_l(t)$, and $S_{pop}(t)=e^{-H_{pop}(t)}$. 
Fig. \ref{fig:survival} shows the population survival curves and population hazards of the GPTCM when assuming two clusters and Weibull distributed survival. 
It is interesting that the population hazard function of the GPTCM can be multimodal, because $h_{pop}(t|X)$ is a mixture of $L$ density functions, which is beyond the basic shapes of the hazard function [e.g. constant, decreasing, increasing, unimodal (up-then-down), or bathtub (down-then-up) shape] \citep{Christen2025}. 



The importance measure of cluster-specific survival can provide valuable insight for developing effective strategies to improve or intervene the entire system, applicable to both biomedical applications and reliability engineering. Such measures help to identify which clusters should receive attention in survival improvement efforts.
The population survival function at a given time $t$ is expressed as a function of the $L$ clusters' survival at that time, i.e.
$$
S_{pop}(t) = f(S_1(t),S_2(t),...,S_L(t)).
$$
The Birnbaum measure \citep{Birnbaum1969} can be used to evaluate the survival importance of different clusters 
given by
$$
\frac{\partial S_{pop}(t)} {\partial S_l(t)} =
S_{pop}(t) \theta p_l,\ l=1,...,L.
$$
Note that the Birnbaum measure does not account for any time-dependent covariate. 
A systematic overview of different importance measures can be found in \cite{Wu2022}. 


\section{Simulation study}\label{sec:simulation}

We provide insights about the parameter estimation of the proposed GPTCM in Section \ref{subsec:formulation} by using Monte Carlo simulations. 
We consider sample sizes of $n=200, \; 500$ and $1000$. 
Each sample/patient has two clinical covariates (i.e. one row of the clinical data matrix $\mathbf X_0\in \mathbb R^{n\times 2}$), and has cells belonging to $L=3$ tumor cell subtypes with each subtype consisting of two cell-type-specific covariates (i.e. one row of data matrix $\mathbf X_l\in \mathbb R^{n\times 2}$, $l\in \{1,...,L\}$). 
Each sample also has tumor cell subtype proportions data (i.e. one row of the proportions data matrix $\mathbf p\in [0,1]^{n\times 2}$). 
Every covariate is generated independently from the standard normal distribution except the first clinical variable generated from the Bernoulli distribution. 
The tumor cell subtype proportions of each sample is generated independent from the Dirichlet distribution. 
The survival times are generated based on the population survival function (\ref{S_pop_new}) using rate parameter $\theta=\exp(\xi_0+\mathbf X_0\bm\xi)$, 
and using the Weibull distributed survival functions with 
mean parameters $\log \bm\mu_{l}=\mathbf X_{l} \bm \beta_l$, $l\in \{1,...,L\}$. 
Censoring is generated through an exponential distribution with approximately $50\%$ censoring rate. 
The true values of all parameters are shown in Table \ref{table_mle}. 

The maximum likelihood (ML) estimation is to maximize the log-likelihood function for right-censored survival data, i.e. 
$$ 
\mathcal L(\bm\vartheta | \mathcal D) = \prod_{i=1}^{n} f_{pop}(t_i|\mathbf X_{0}, \mathbf X_{1},\mathbf X_2,\mathbf X_3, \mathbf p)^{\delta_i} S_{pop}(t_i|\mathbf X_{0}, \mathbf X_{1},\mathbf X_2,\mathbf X_3, \mathbf p)^{{1-\delta_i}},
$$ 
where $\bm\vartheta$ consists of all unknown parameters and $\mathcal D = \{t_i, \delta_i, \mathbf X_{0}, \mathbf X_{1},\mathbf X_2,\mathbf X_3, \mathbf p\}_{i=1}^n$ consists of all data information including each sample's observed survival time $t_i$, censoring indicator $\delta_i$, covariates, and cell subtype proportions. 
The R function \texttt{nlminb} using the adaptive nonlinear least-squares algorithm is to perform the optimization. 
We repeat the scenario of each sample size $1000$ times to obtain the ML estimates with mean and standard error. 

Table \ref{table_mle} shows that the mean squared error (MSE) of each estimate decreases with the increase of sample size as we expected. 
The performance of the ML estimates of all parameters, except for the Weibull's shape parameter $\log(\kappa)$, are close to their true values. 
However, future work for further investigations with more simulation scenarios (e.g. more covariates and model misspecification) and applications to real data is needed to better understand the implications of the proposed model. 

\begin{table}[h!]
\centering
\caption{Simulation results with maximum likelihood estimates (standard errors in parentheses) and the mean squared errors for different sample sizes}
\label{table_mle}
\begin{tabular}{crrc c rc c rc}
  \hline
  \hline
 Parameter & Truth & \multicolumn{1}{c}{Estimate} & MSE && \multicolumn{1}{c}{Estimate} & MSE && \multicolumn{1}{c}{Estimate} & MSE\\ 
  \hline
  &  & \multicolumn{2}{c}{$n=200$} &&  \multicolumn{2}{c}{$n=500$} && \multicolumn{2}{c}{$n=1000$} \\ 
  \cmidrule(r){3-4}
  \cmidrule(r){6-7}
  \cmidrule(r){9-10}
$\log(\kappa)$ & 1.10 & 1.69 (0.098) & 0.361 && 1.64 (0.060) & 0.293 && 1.62 (0.041) & 0.271 \\ 
$\xi_1$ & -0.80 & -0.81 (0.193) & 0.037 &&-0.80 (0.120) & 0.014 && -0.79 (0.082) & 0.007\\
$\xi_2$ &  0.90 & 1.05 (0.242) & 0.080 && 1.05 (0.150) & 0.044 && 1.03 (0.100) & 0.028\\
$\xi_3$ &  0.60 &0.71 (0.134) & 0.031 && 0.71 (0.080) & 0.020 && 0.71 (0.054) & 0.015\\
$\beta_{11}$ & 0.40 & 0.29 (0.086) & 0.018 && 0.29 (0.049) & 0.014 && 0.29 (0.034) & 0.013\\
$\beta_{12}$ & -0.30 & -0.22 (0.084) & 0.013 && -0.22 (0.049) & 0.009 && -0.22 (0.034) & 0.008\\
$\beta_{21}$ & 0.25 & 0.18 (0.076) & 0.010 && 0.18 (0.046) &0.007 &&0.18 (0.032) & 0.005\\
$\beta_{22}$ & -0.45 & -0.33 (0.084) & 0.021 && -0.33 (0.051) &0.017 && -0.33 (0.036) & 0.016\\
$\beta_{31}$ & -0.20 & -0.15 (0.075) & 0.008 && -0.15 (0.048) & 0.005 &&-0.15 (0.033) & 0.004\\
$\beta_{32}$ & 0.30 & 0.22 (0.080) & 0.012 && 0.22 (0.050) & 0.008 &&0.22 (0.035) & 0.007\\
  \hline
  \hline
\end{tabular}
\end{table}

\section{Conclusion}\label{sec:conclusion}

We have presented a new promotion time cure model GPTCM that is a generalized version of the classical PTCM. 
The new formulation consists of the part of a mixture of survival distributions that is strongly motivated by biological intra-tumor heterogeneity of patients rather than mathematical construction of a mixture model. 
The new modeling framework is flexible to model survival data with intra-sample heterogeneity in biomedicine or intra-system heterogeneity in reliability engineering. 
Note that both the PTCM and GPTCM assume the latent variable $N$ (i.e. number of clonogenic cells) independent of time $t$. 
\cite{Cha2018} presented various shock models, including the PTCM as a special case, from the counting process of view. 
Therefore, a future direction for an extension of the GPTCM is to model the dynamics of the number of clonogenic cells by treating $N(t)$ as a counting process, which can better mimic tumor evolution.

\section*{Acknowledgments}

This work was supported by the University of Oslo innovation funds, ERA PerMed under the ERA-NET Cofund scheme of the European Union's Horizon 2020 research and innovation framework program (grant ‘SYMMETRY’ ERAPERMED2021-330). 
The authors would like to thank Manuela Zucknick for discussions.

\appendix

\section{Mixture survival model}
\label{sec:appendixA}

When we assume only one tumor cell left active after an initial treatment, this tumor cell belongs to one of the $L$ tumor cell subtypes, with probability $\mathbb P(N_l=1)=p_l$, $l\in\{1,2,...,L\}$, $\sum_{l=1}^L p_l=1$. 
Using the same notations as Section \ref{subsec:formulation}, now the population survival function is
{\footnotesize
\begin{align*}
S_{pop}(t) &= \mathbb P(N=0) + \mathbb P(\bm W_1 >t,... \bm W_L>t, N=\sum_{l=1}^L N_l= 1 )\\
&= \mathbb P(N=0) + \sum_{l=1}^L\mathbb P(\bm W_l >t | N_l = 1 ) \cdot \mathbb P(N_l=1) \\
&= \mathbb P(N=0) + \sum_{l=1}^Lp_lS_l(t).
\end{align*}
}
This is a classical mixture model. However, the assumption with only one tumor cell left active after an initial treatment is not biologically meaningful in cancer research.

\section{Last-activation scheme}
\label{sec:appendixB}

Our approach in Section \ref{subsec:formulation} can be adapted straightforwardly for the last-activation scheme. 
The population survival function is now

{\small
\begin{align*}
  \tilde S_{pop}(t) &= \mathbb P(N=0) + 
  \mathbb P\left(\max\{\max\{\bm W_1\},...,\max\{\bm W_L\}\} >t, N>0 \right) \\
  &= \mathbb P(N=0) + \mathbb P\left(\max\{\max\{\bm W_1\},...,\max\{\bm W_L\}\} > t| N>0 \right) \mathbb P(N>0)\\
  &= \mathbb P(N=0) + \left[1 - \mathbb P\left(\max\{\max\{\bm W_1\},...,\max\{\bm W_L\}\} \le t| N>0 \right)\right] \mathbb P(N>0)\\
  &= \mathbb P(N=0) + \mathbb P(N>0) - \mathbb P\left(\max\{\max\{\bm W_1\},...,\max\{\bm W_L\}\} \le t| N>0 \right) \mathbb P(N>0)\\
  &= 1 - \mathbb P(\max\{\bm W_1\} \le t, ..., \max\{\bm W_L\} \le t, N>0 ) \\
  &= 1- \sum_{k=1}^\infty\mathbb P\left(\max\{\bm W_1\} \le t, ..., \max\{\bm W_L\}\le t | N=k\right) \cdot \mathbb P(N=k)\\
  &= 1 - \sum_{k=1}^\infty\sum_{\substack{\text{All config.}\\(N_1,...,N_L)\\ \text{ with sum }k}} \mathbb P\left(\max\{\bm W_1\} \le t, ..., \max\{\bm W_L\} \le t |N=k, N_1,...,N_L\right) 
  \cdot \mathbb P(N=k,N_1,...,N_L).
\end{align*}
}
Denote the component-specific cdf as $F_l(t)=1-S_l(t)$. By using the multinomial-Poisson transformation, the multinomial theorem and the power series of the exponential function, we obtain 
{\small
\begin{align*}
  \tilde S_{pop}(t) 
  &= 1 - 
  \sum_{k=1}^\infty\sum_{\substack{\text{All config.}\\(N_1,...,N_L)\\ \text{ with sum }k}} \prod_{l=1}^L \mathbb P(\max\{\bm W_l\} \le t | N_l) \cdot \mathcal Pois(k;\theta)\mathcal Mult(k;\bm p)\\ 
  &= 1 - \sum_{k=1}^\infty\sum_{\substack{\text{All config.}\\(N_1,...,N_L)\\ \text{ with sum }k}} \prod_{l=1}^L F_l(t)^{N_l} \cdot \mathcal Pois(k;\theta)\mathcal Mult(k;\bm p)\\ 
  &= 1 - \left\{e^{\theta\sum_{l=1}^Lp_lF_l(t)} - 1 \right\} e^{-\theta} \\
  &= 1 + e^{-\theta} - e^{-\theta\{1-\sum_{l=1}^Lp_lF_l(t)\}}\\
  &= 1 + e^{-\theta} - e^{-\theta\sum_{l=1}^Lp_lS_l(t)}.
\end{align*}
}

 \bibliographystyle{apalike}
 \bibliography{refs}





\end{document}